\documentclass[showpacs,showkeywords,superscriptaddress, preprint,aps]{revtex4}

\usepackage{graphicx}
\usepackage{dcolumn}
\usepackage{bm}

\begin{document}

\title{Local Lattice Instability and Superconductivity in
La$_{1.85}$Sr$_{0.15}$Cu$_{1-x}$M$_x$O$_4$ (M=Mn, Ni, and Co)}

\author{C. J. Zhang}
\affiliation{National Institute of Advanced Industrial Science and
Technology, 1-1-1 Umezono, Tsukuba 305-8568, Japan}

\author{H. Oyanagi}
\thanks{Author to whom correspondence should be addressed; h.oyanagi@aist.go.jp}
\affiliation{National Institute of Advanced Industrial Science and
Technology, 1-1-1 Umezono, Tsukuba 305-8568, Japan}


\begin{abstract}
Local lattice structures of
La$_{1.85}$Sr$_{0.15}$Cu$_{1-x}$M$_x$O$_4$ (M=Mn, Ni, and Co) single
crystals are investigated by polarized extended x-ray absorption
fine structure (EXAFS). The local lattice instability at low
temperature is described by in-plane Cu-O bond splitting. We find
that substitution of Mn for Cu causes little perturbation of local
lattice instability while Ni and Co substitution strongly suppresses
the instability. The suppression of superconductivity by Cu-site
substitution is related to the perturbation of lattice instability,
indicating that local lattice instability (polaron) plays an
important role in superconductivity.
\end{abstract}
 \maketitle

\bigskip

\section{Introduction}
In recent years there has been a growing interest in the possibility
that the metallic phase of cuprate superconductors has an
instability towards microscopically phase-separated charge and spin
inhomogeneities \cite{1} and that this instability might be related
to the high temperature superconductivity (HTSC) itself \cite{2}. In
an inhomogeneity picture, the modulated charges (stripes) could in
principle support either superconductivity or charge density wave
\cite{3}. When the stripes are static, charge density wave
overwhelms. However, if the stripes are dynamic, the charge
fluctuations might suppress the charge density wave and open a
pseudogap that serves as a precursor for superconductivity. This
nanometer-scale phase instability could lead to a strong
spin-charge-phonon coupling, giving possible evidence of
unconventional electron-phonon coupling in HTSC.

The existence of static stripes is now generally accepted due to the
observation of inhomogeneous charge distribution in certain
compounds such as La$_{1.6-x}$Nd$_{0.4}$Sr$_x$CuO$_4$ and
La$_{1.875}$Ba$_{0.125}$CuO$_4$ \cite{4,5}. However, dynamic
stripes---which are particularly important---are not easily and
unambiguously detectable. Due to the absence of a clear experimental
signature, the possible existence of dynamic inhomogeneous state has
been difficult to establish \cite{6}. So far, dynamic inhomogeneity
has been suggested by the observation of a second order parameter
well above $T_c$. For example, a multi-component response, as a sign
of dynamic inhomogeneous state in La$_{2-x}$Sr$_x$CuO$_4$ has been
consistently observed in different experiments, such as muon-spin
rotation, scanning tunneling spectroscopy, and angle resolved
photoemission spectroscopy \cite{7,8,9}. In a recent inelastic
scattering study, Reznik $et$ $al$. found a clearly discernible
characteristic anomaly in the spectrum of phonons in optimally-doped
La$_{1.85}$Sr$_{0.15}$CuO$_4$ \cite{10}. This anomaly indicates the
resonance between phonons and the motion of ions that form the
copper oxide lattice, reflecting the possible existence of dynamic
charge inhomogeneity in optimally-doped
La$_{1.85}$Sr$_{0.15}$CuO$_4$. However, inelastic scattering
measurements do not yield direct information on ``modulated" Cu-O
bond distances. Clearly, if the charge is inhomogeneously
distributed in the CuO$_2$ plane, such that some copper sites have
more charge than others, a modulation of in-plane Cu-O bond length
would occur. Thus, a high-precision measurement of the in-plane Cu-O
bond length ``distribution" would reveal the possible existence of
an inhomogeneous state. In this paper we present an accurate
determination of the in-plane Cu-O bond length modulation by
detecting the Cu-O atomic displacements. Splitting of the in-plane
Cu-O bond distance below a characteristic temperature $T^*$ is
clearly present, providing a solid evidence for the existence of
local lattice instability in optimally-doped
La$_{2-x}$Sr$_x$CuO$_4$. We also report specific perturbations of
local lattice instability by doping with different impurities (Mn,
Ni, and Co), which is related to the different behaviors on the
suppression of superconductivity. The results strongly suggest that
the local lattice instability plays an important role in HTSC.

\section{Experiment}

Single crystal samples of La$_{1.85}$Sr$_{0.15}$Cu$_{1-x}$M$_x$O$_4$
(M=Mn, Ni, Co) were grown by the traveling solvent floating zone
method \cite{11}. For the feed rod, stoichiometric polycrystalline
powders of La$_{1.85}$Sr$_{0.15}$Cu$_{1-x}$M$_x$O$_4$ (M=Mn, Ni, Co)
were preheated twice at 1150$^{\circ}$C. X-ray diffraction
measurements were performed on these polycrystalline powders in
order to confirm that the samples were single phase. Then, an
appropriate amount of extra CuO was added into the polycrystalline
powder in order to compensate for the loss of Cu during the crystal
growth and the mixture was thoroughly ground. The obtained fine
powders of each sample were placed into a thin-walled rubber tube
and formed into a cylindrical seed rod under hydrostatic pressure.
The seed rods were sintered at 1250$^{\circ}$C for 24 h in air. The
solvent material with the composition of
La:Sr:Cu:M=1.85:0.15:(4-$x$):$x$ for each $x$ was prepared. We used
0.3 g of the solvent and a La$_{1.85}$Sr$_{0.15}$CuO$_4$ single
crystal as a seed rod. The single crystal growth rate of Mn-doped
samples was kept constant at 0.2 mm/h and the growth rate of
Ni-doped and Co-doped samples was 0.5 mm/h. In order to eliminate
oxygen deficiencies, all the crystals were annealed under oxygen gas
flow at 900$^{\circ}$C for 50 h, cooled to 500$^{\circ}$C at a rate
of 10$^{\circ}$C/h, and annealed at 500$^{\circ}$C for 50 h. The
superconducting transition temperature of the samples was determined
by a SQUID magnetometer (Quantum Design, MPMS). The applied magnetic
field is 10 Oe with field-cooling. The magnetic field is applied
perpendicular to the CuO$_2$ plane.

EXAFS measurements were performed at BL13B at Photon Factory,
Tsukuba. The energy and maximum electron current were 2.5 GeV and
400-500 mA, respectively. A directly water-cooled silicon (111)
double-crystal monochromator was used, covering the energy range of
4-25 keV. The energy of the incident x-ray was calibrated by
assigning the first shoulder of the Cu foil spectrum (Cu $K$-edge)
to 8.9803 keV. The energy resolution was better than 2 eV at 9 keV.
The drift in the energy calibration was below 5\%\ of the energy
resolution. The number of the counted photons is $\sim$10$^8$
photons/s. The sample was mounted in a closed-cycle He refrigerator
and the temperature was monitored with an accuracy of $\pm$0.1 K.
The sample sizes were approximately 3$\times$3$\times$0.8 mm$^3$.
The measurements of the in-plane polarized EXAFS spectroscopies were
made using polarization of the light parallel to the CuO$_2$ plane.
For the measurements of the out-of-plane polarized spectroscopies,
we used a different sample holder so that the electric field of the
polarized light is parallel to the $c$-axis of the sample. The
details of the experimental setup and data acquisition process have
been described elsewhere \cite{12}. Here we used a state-of-the-art
germanium 100-pixel array detector (PAD) collecting signals over a
segmented solid angle using a grazing-incidence geometry.

\section{Results and discussion}

\subsection{Local lattice instability in La$_{1.85}$Sr$_{0.15}$CuO$_4$}

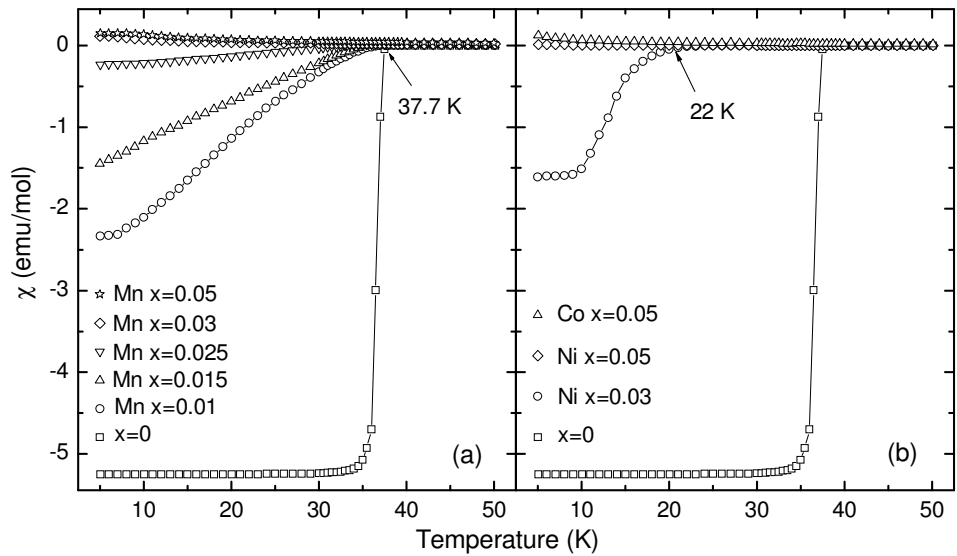
Figure 1 gives the temperature dependence of magnetic susceptibility
for La$_{1.85}$Sr$_{0.15}$Cu$_{1-x}$M$_x$O$_4$ (M=Mn, Ni, Co)
samples (Meissner signals). The onset superconducting transition
temperature ($T_c^{onset}$) for La$_{1.85}$Sr$_{0.15}$CuO$_4$ is
37.7 K. Superconducting transition is observed in $x$$\leq$0.025
samples of La$_{1.85}$Sr$_{0.15}$Cu$_{1-x}$Mn$_x$O$_4$. Strikingly,
$T_c^{onset}$ remains constant at about 36.5 K in Mn-doped samples,
while it decreases rapidly in Ni- and Co-doped samples. For example,
the $T_c^{onset}$ value of
La$_{1.85}$Sr$_{0.15}$Cu$_{1-x}$Ni$_x$O$_4$ with $x$=0.03 is about
22 K. In both Mn- and Ni-doped samples, the superconducting volume
fraction decreases with increasing doping content.

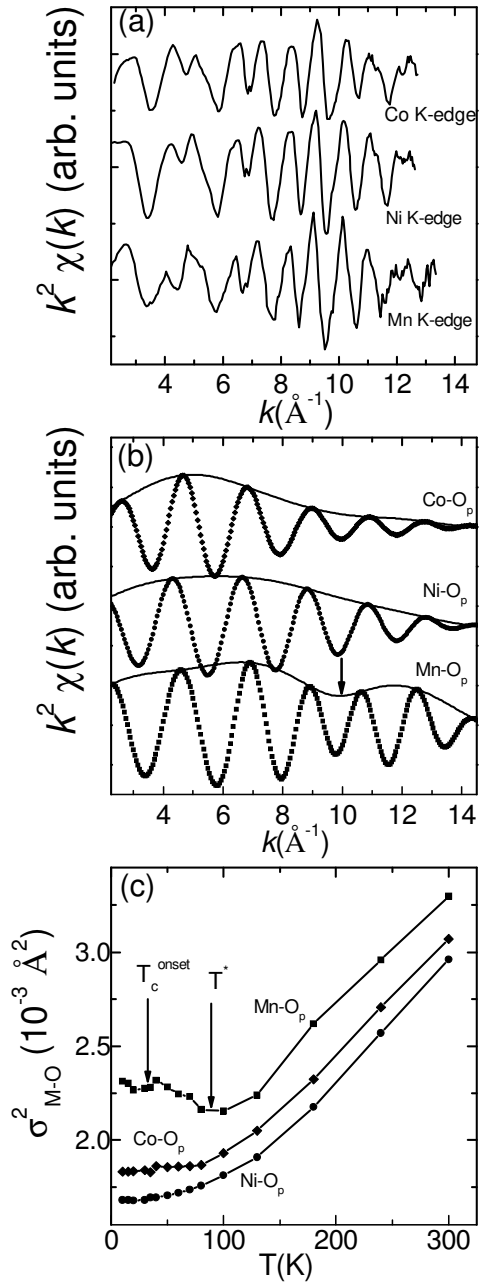
Figures 2(a) and (b) show representative examples of the in-plane
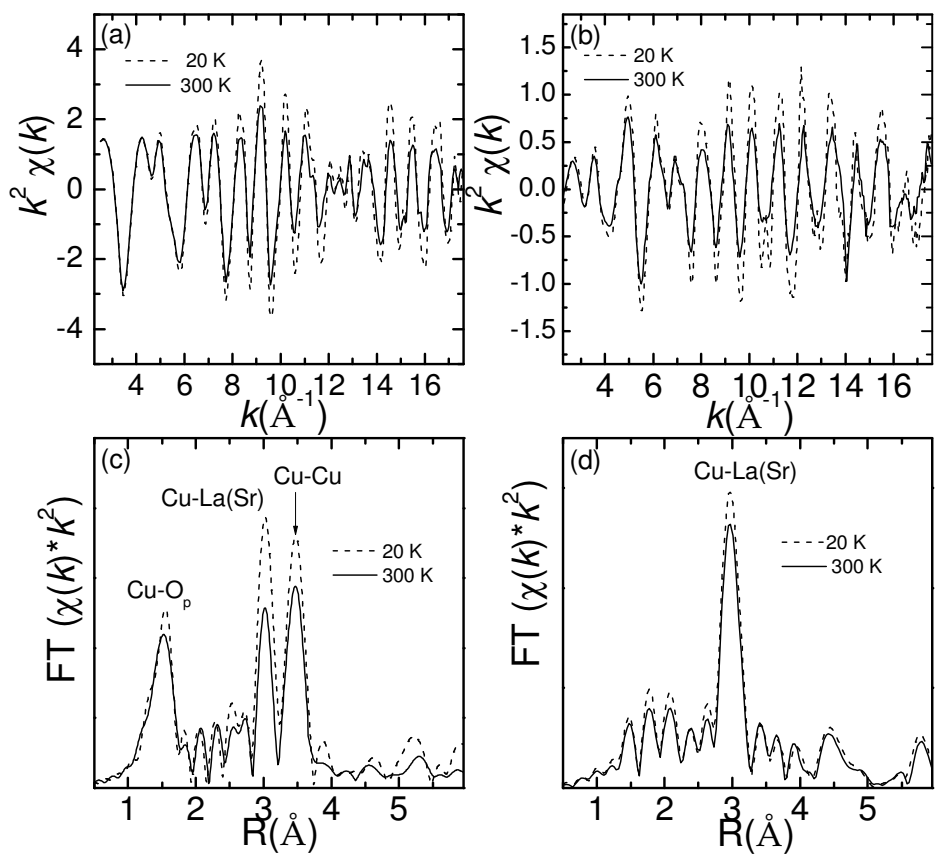
and out-of-plane polarized Cu $K$-edge EXAFS oscillations (weighted
by $k^2$) of La$_{1.85}$Sr$_{0.15}$CuO$_4$, respectively. In both
polarization geometries, the EXAFS oscillations taken at 300 K and
20 K are shown. The magnitude of the EXAFS oscillations is enhanced
at low temperature. Figures 2(c) and (d) show the complex Fourier
transform (FT) magnitude function of the EXAFS oscillations for
La$_{1.85}$Sr$_{0.15}$CuO$_4$ with the in-plane and out-of-plane
configurations, respectively. The complex FT magnitude function is
related to the atomic radial distribution function around the
central Cu atom. Compared to the real crystallographic distances,
the coordination shell peaks in Figs. 2(c) and (d) are shifted to
lower $R$ due to scattering phase shifts. In the in-plane
configuration, prominent peaks at radii $R$$\sim$1.5 \AA,
$R$$\sim$3.0 \AA, and $R$$\sim$3.5 \AA\ are the Cu-O, Cu-La, and
Cu-Cu correlations. In the out-of-plane configuration, the peaks
corresponding to the in-plane Cu-O and Cu-Cu correlations are
significantly damped. The prominent peak at radius $R$$\sim$3.0 \AA\
corresponds to the Cu-La correlation. There are several peaks in the
1.3$<$$R$$<$2.2 \AA\ range where the out-of-plane Cu-O peak appears.

The experimental EXAFS, $\chi$($k$), is analyzed by using of the
IFEFFIT analysis package \cite{13}. The fitting of EXAFS data is
performed for each orientation by constraining the structural
parameters of those paths that contribute to individual
polarization. In the in-plane polarization, we fit the experimental
data in the 1.0$<$$R$$<$2.0 \AA\ range. Except the radial distance
$R$ and the mean-square relative displacement (MSRD) $\sigma$$^2$,
all other parameters are kept constant in the conventional least
squares paradigm. The dashed line in Fig. 3(a) represents a typical
curve fitted to the experimental result at 300 K. It can be seen
that this curve shows a good agreement with the experimental curve
(solid line) in the 1.0$<$$R$$<$2.0 \AA\ range. For the out-of-plane
geometry, we fit the experimental data in the 1.2$<$$R$$<$3.8 \AA\
range, taking into account the contribution from all possible
scattering paths. A typical fitted curve is given as the dashed line
in Fig. 3(b). In Figs. 3(c) and (d) we compare the fitted curves and
experimental data in $k$-space for the in-plane configuration and
the out-of-plane configuration, respectively. In both
configurations, the fitting curves reproduce the experimental data
well.

Figure 4 shows the temperature dependence of MSRD of the in-plane
Cu-O bond, $\sigma$$^2_{Cu-O_p}$ and the out-of-plane Cu-O bond,
$\sigma$$^2_{Cu-O_{ap}}$ for La$_{1.85}$Sr$_{0.15}$CuO$_4$. In the
high temperature region, $\sigma$$^2_{Cu-O_p}$ decreases with
decreasing temperature. In contrast, below a characteristic
temperature $T^*$ ($T^*$$\sim$80 K), it exhibits a remarkable
upturn, indicating the occurrence of local lattice distortion
\cite{14,15,16,17}. Previously, a similar local lattice distortion
was found and explained by using a two-componen model where a
distorted local low temperature tetragonal (LTT) lattice coexists
with the undistorted local low temperature orthorhombic (LTO)
lattice \cite{14}. In this study, we systematically analyzed the
polarized Cu $K$-edge EXAFS data along the $c$-axis. In Fig. 4 we
also plot the temperature dependence of MSRD of the out-of-plane
Cu-O bond $\sigma$$^2_{Cu-O_{ap}}$ for
La$_{1.85}$Sr$_{0.15}$CuO$_4$. We do not observe any anomaly in the
temperature dependence of MSRD of the out-of-plane Cu-O bond. This
result suggests that the out-of-plane Cu-O bond is not involved in
the local lattice instability, i.e., the local lattice distortion
occurs within the CuO$_2$ plane.

Closer examination of the EXAFS oscillations reveals the detailed
temperature dependence of the in-plane Cu-O bond distortion. Figure
5 plots the Fourier-filtered (back-transforming over 0.9$<$$R$$<$2.0
\AA) EXAFS oscillations and amplitudes of the in-plane Cu-O bond
from 300 K to 10 K. It is found that the oscillations are very
similar up to $k$$\sim$12 \AA$^{-1}$ at all temperatures. However,
for $T$$<$80 K the local minimum in the amplitude and the
irregularity in the phase near 12.5 \AA$^{-1}$ constitute a ``beat",
which signifies the presence of at least two Cu-O shells having
different bond distances. Using the relation $\Delta$$R$=$\pi$/2$k$
between the separation of the two shells ($\Delta$$R$) and the
position of the beat ($k$), the Cu-O distances are determined to
differ by $\sim$0.12 \AA. Taking into account the average Cu-O bond
distance of $\sim$1.88 \AA, the long and short Cu-O bond distances
are determined to be $\sim$1.94 \AA\ and 1.82 \AA, respectively.
That means, some oxygens are shifted towards or away from their
adjacent Cu sites by $\sim$0.06 \AA.

The local lattice instability below $T^*$$\sim$ 80 K is described by
the in-plane Cu-O bond splitting. Due to the well-known fact that
there is no static lattice/charge modulation in optimally-doped
La$_{1.85}$Sr$_{0.15}$CuO$_4$ and the failure to detect the
inhomogeneous state by any slow techniques, we suggest that the
local lattice instability occurs ``dynamically". The dynamic lattice
fluctuation time is below the picosecond (10$^{-12}$ second) range,
which can be detected only by certain ``fast" probes, such as EXAFS
and neutron diffraction \cite{6,18}. The powerful fast probes which
\emph{freeze} the motion of ions on short timescales
($\sim$10$^{-15}$ second), e.g. EXAFS, are eminently suitable for
detection of these fluctuation patterns \cite{18}. Lattice
instability at this length scale is expected to produce exotic
lattice dynamical properties such as a certain kind of phonon mode
softening and strong coupling between the lattice and charge degrees
of freedom, leading to polaron formation. Regarding the specific
nature of the polaron, two different types of Cu-O bond stretching
modes can be considered as possible candidates: the
pseudo-Jahn-Teller mode \cite{19,20}; and the $Q_2$-type mode
\cite{21,22}. These modes are schematized in Fig. 6. Below $T_c$,
the beat feature is weakened but it is still clearly discernible,
which indicates the persistence of dynamic lattice instability below
$T_c$. Incidentally the sharp decrease of the local Cu-O lattice
displacements was found in ion channeling experiments, where a drop
of the variable amplitude at $T_c^{onset}$ was detected \cite{23}.
These results indicate that the vibration of planar oxygen becomes
silent as the superconducting transition occurs. We conclude that
the decrease in vibration is due to the transition of the CuO$_2$
planar lattice dynamics from an incoherent state to a coherent state
as the pairing occurs.

\subsection{Perturbations of local lattice instability by Mn, Ni, and Co doping}

Besides the characterization of the in-plane Cu-O bond local lattice
instability in La$_{1.85}$Sr$_{0.15}$CuO$_4$, another important
topic is whether or not and how this local lattice instability
relates to the nature of the superconductivity in this system. In
the following we investigate some specific perturbations of the
local lattice instability by impurity doping. We have used three
different transition metal elements (Mn, Ni, and Co) to replace the
Cu in the La$_{1.85}$Sr$_{0.15}$CuO$_4$ parent compound. For all the
doped La$_{1.85}$Sr$_{0.15}$Cu$_{1-x}$M$_x$O$_4$ (M=Mn, Ni, Co)
single crystal samples, polarized Cu $K$-edge EXAFS experiments have
been performed from 300 K to 5 K, and the same data analysis
procedure has been performed. To simplify the data, we plot the
temperature dependence of $\sigma$$^2_{Cu-O_p}$ for
La$_{1.85}$Sr$_{0.15}$Cu$_{1-x}$M$_x$O$_4$ (M=Mn, Ni, Co) samples in
Figs. 7(a) and (b). It is found that the doping of Mn, Ni, and Co
leads to different perturbations of the local lattice instability.
That is, the introduction of Ni and Co dopants at the Cu site
strongly depresses the upturn of $\sigma$$^2_{Cu-O_p}$ below $T^*$.
With 5\% of Ni or Co doping the upturn behavior is completely
disappeared. In contrast, the Mn dopant has less perturbation on the
upturn behavior. We notice that the upturn of $\sigma$$^2_{Cu-O_p}$
is still significant in the 5\% Mn-doped sample. In order to see the
difference in the perturbation clearly, we fit the temperature
dependence of $\sigma$$^2_{Cu-O_p}$ for
La$_{1.85}$Sr$_{0.15}$CuO$_4$ by using the correlated Einstein model
\cite{24}, via the following equation:

\begin{eqnarray}
\sigma_{th}^2(T)=\frac{\hbar^2}{2\mu
k_B\Theta_E}\text{coth}(\frac{\Theta_E}{2T}),
\end{eqnarray}

\noindent{where $\Theta$$_E$ is the Einstein temperature and $\mu$
is the reduced mass of the pair of atoms involved in the bond. The
calculated $\sigma$$_{th}^2$($T$) curve is shown as the dashed line
in Fig. 7. One can clearly see a large deviation between the
measured and calculated $\sigma$$^2_{Cu-O_p}$ below 80 K, due to the
occurrence of local lattice instability.}

We find that all the dopants (Mn, Ni, and Co) lead to an increase in
$\sigma$$^2_{Cu-O_p}$, which is likely due to local lattice
rearrangement around the dopants. In this case, the
$\sigma$$^2_{Cu-O_p}$ could be given as a superposition of a
temperature-independent term ($\sigma$$^2_{imp}$) and a
temperature-dependent dynamic term. For these samples, we also fit
the experimental $\sigma$$_{Cu-O_p}^2$($T$) curves using a simple
equation:

\begin{eqnarray}
\sigma^2_{Cu-O_p}(T)=\sigma^2_{th}(T)+\sigma^2_{imp}.
\end{eqnarray}

\noindent{The fitting results are shown as the dashed curves in Fig.
7.}

The detailed variation of $\sigma$$^2_{Cu-O_p}$ can be seen more
clearly in Figs. 8(a) and (b) where the $\sigma$$^2_{Cu-O_p}$ values
below 100 K are shown. In all Mn-doped samples, the upturn of
$\sigma$$^2_{Cu-O_p}$ occurs below $T^*$$\sim$80 K. In Ni-doped
samples, the $T^*$ slightly decreases. For example, the $T^*$ value
is about 70 K in 3\% Ni-doped sample. It is interesting to notice
that the drop of $\sigma$$^2_{Cu-O_p}$ occurs at $T_c^{onset}$. Thus
it is clear that the upturn of $\sigma$$^2_{Cu-O_p}$ below $T^*$ is
due to the occurrence of local lattice instability while the drop of
$\sigma$$^2_{Cu-O_p}$ below $T_c^{onset}$ is due to the
superconducting phase coherence. Using two simple equations we can
roughly estimate the magnitudes of the upturn part and the drop
part:

\begin{eqnarray}
\text{For}\ T_c<T<T^*:
\sigma^2_{Cu-O_p}(T)=\sigma^2_{th}(T)+\sigma^2_{imp}+\sigma^2_{ins}(T),
\end{eqnarray}

\begin{eqnarray}
\text{For}\ T<T_c:
\sigma^2_{Cu-O_p}(T)=\sigma^2_{th}(T)+\sigma^2_{imp}+
\sigma^2_{ins}(T)-\sigma^2_{coh}(T),
\end{eqnarray}

\noindent{where $\sigma$$^2_{ins}$ represents the contribution of
local lattice instability to the total $\sigma$$^2_{Cu-O_p}$, and
$\sigma$$^2_{coh}$ is due to the phase coherence below $T_c$. Using
Eqs. (3) and (4), we can estimate the magnitudes of
$\sigma$$^2_{ins}$ and $\sigma$$^2_{coh}$ for each sample by
subtracting the theoretically modeled Debye-Waller factor from the
experimental data. The results are shown in Fig. 8(c) and (d).}

From Fig. 8(c) one can see that $\sigma$$^2_{ins}$ has the largest
magnitude in the optimum superconductor
La$_{1.85}$Sr$_{0.15}$CuO$_4$. With increasing Mn doping, the
magnitude of $\sigma$$^2_{ins}$ slightly decreases but the
contribution from $\sigma$$^2_{ins}$ persists in all samples up to
$x$=0.05, indicating that Mn doping does not suppress the local
lattice instability. On the other hand, the contribution of
$\sigma$$^2_{ins}$ decreases rapidly in the Ni- and Co-doped
samples. With 5\% of Ni(Co)-doping the upturn of
$\sigma$$^2_{Cu-O_p}$ is completely smeared out, indicating that the
local lattice instabilities are strongly suppressed by Ni and Co
dopants. Fig. 8(d) plots the magnitude of the drop of
$\sigma$$^2_{Cu-O_p}$ due to the superconducting phase coherence
below $T_c^{onset}$. The drop in $\sigma$$^2_{Cu-O_p}$ has the
largest magnitude in the optimally-doped
La$_{1.85}$Sr$_{0.15}$CuO$_4$ sample. In Ni- and Co-doped samples
the magnitude of this drop decreases with increasing doping
concentration and finally the drop disappears in non-superconducting
samples. It is interesting to notice that in Mn-doped samples, the
drop in $\sigma$$^2_{Cu-O_p}$ is observed in all samples, despite
the decrease in magnitude.

On the basis of these results we can draw the conclusion that the
Mn-doped samples favor the local lattice instability while the Ni
and Co dopants are strongly harmful to the local lattice
instability. The perturbation of local lattice instability is
related to the suppression of superconductivity induced by impurity
doping. That is, the introduction of Ni and Co dopants strongly
suppresses the local lattice instability and thus severely
suppresses the superconducting transition. In contrast to the Ni and
Co dopants, the Mn dopants cause less perturbation on the local
lattice instability, resulting in a constant onset superconducting
transition temperature. It is reasonable to imagine that in Mn-doped
samples, the introduction of Mn dopants leads to the separation of
the superconducting domains. Near the Mn dopants, regions of
non-superconducting domains are formed. With increasing Mn doping,
more and more non-superconducting regions are formed and large
superconducting domains could be divided into small superconducting
droplets. As the distance between the superconducting droplets
becomes large, the tunneling is prohibited and the superconductivity
disappears.

\subsection{Local lattice structure around Mn, Ni, and Co dopants}

We also investigate the local lattice structures around the Mn, Ni,
and Co dopants by analyzing the Mn, Ni, and Co $K$-edges EXAFS data.
Figure 9(a) shows examples of the in-plane EXAFS oscillations
(weighted by $k^2$) for La$_{1.85}$Sr$_{0.15}$Cu$_{1-x}$M$_x$O$_4$
with $x$=0.05 at 40 K at the Mn, Ni, and Co $K$-edges. Figure 9(b)
gives the Fourier-filtered EXAFS oscillations and amplitudes of the
in-plane M-O (M=Mn, Ni, Co) bond at 40 K. A distinct difference can
be seen between the Fourier-filtered EXAFS oscillations of the Ni(or
Co)-O bond and that of the Mn-O bond. That is, a ``beat" feature is
clearly present at $k$$\sim$10 \AA\ for the Mn-O bond oscillation,
which signifies the presence of two in-plane Mn-O bonds with
different distances. The two Mn-O distances are estimated to differ
by $\sim$0.16 \AA. The ``beat" feature is indiscernible for the
Ni(or Co)-O bond case. The temperature dependence of in-plane M-O
(M=Mn, Ni, and Co) bond MSRD $\sigma$$^2_{M-O_p}$ is shown in Fig.
9(c). The upturn of $\sigma$$^2_{Mn-O_p}$ below $\sim$100 K is
consistent with the existence of the two Mn-O bond distances. The
gradual decrease of $\sigma$$^2_{Ni-O_p}$ and $\sigma$$^2_{Co-O_p}$
with decreasing temperature indicates the absence of local lattice
instability around Ni and Co atoms. The Mn, Ni, and Co $K$-edge
EXAFS data constitute direct information on the local lattice
structures around these dopants. For the Mn dopants, the ``beat"
feature and an upturn of $\sigma$$^2_{Mn-O_p}$ appear at low
temperature ($<$100 K), indicating the occurrence of local lattice
instability around the Mn dopants. By contrast, no evidence of local
lattice instability is observed in Ni- and Co-doped samples. These
results suggest that the Mn dopants do not suppress local lattice
instability while the Ni and Co dopants strongly suppress the
instability.

We also find that the substitution of Mn, Ni and Co at the Cu site
leads to a noticeable change in the in-plane Cu-O and M-O (M=Mn, Ni,
and Co) bond distances. In Table I we list the typical bond
distances for these samples at 300 K. It is found that Mn doping
leads to a monotonic increase of the in-plane Cu-O bond distance
while the Ni and Co doping changes the Cu-O bond distance only
slightly. For the M-O bond distance, the average in-plane Mn-O bond
distance is slightly longer than the Cu-O bond distance while the
in-plane Ni-O and Co-O bond distances are comparable to the Cu-O
bond distance.

\subsection{Lattice instability and superconductivity}

We now make some discussion on the local lattice instability and its
interplay with superconductivity. A spatial inhomogeneity model is
schematized in Fig. 10. Below $T^*$, local lattice instability
occurs which leads to phase separation into metallic domains and
insulating ones \cite{1,25,26,27}. If the distance between adjacent
metallic domains ($L$) is shorter than a critical length $\zeta$,
tunneling between the metallic domains is allowed. Phase coherence
is achieved when the quantum tunneling threshold is reached. The
formation of macroscopic superconducting state is through the
spreading of phase coherence by tunneling over the static insulating
domains \cite{28,29,30}. Difference between Mn and Ni (or Co) doping
is due to the fact that Mn atoms conserve the local dynamic domains
while the Ni and Co dopants make the local lattice distortion
``static" because of the strain field. Around the Ni and Co dopants,
the static lattice distortion is temperature-independent, which is
consistent with the term $\sigma$$^2_{imp}$ in Eq. (2). In Ni- and
Co-doped samples, dopants lead to the decrease of dynamic regions.
The decrease of dynamic regions leads to a decrease of
superconducting carrier concentration ($n_s$), resulting in lower
$T_c$ \cite{31}. In this picture, dynamic lattice response plays an
important role in HTSC.

\section{Conclusion}

In conclusion, we present evidence from EXAFS measurements that
dynamic local lattice instability occurs in optimally-doped
La$_{1.85}$Sr$_{0.15}$CuO$_4$. The lattice instability occurs only
within the CuO$_2$ plane while it does not involve the apical
oxygens. Through the substitution of Mn, Ni, and Co for Cu we find
that dynamic lattice instability is intimately related to
superconductivity. While Ni and Co suppress the dynamic instability
and quench superconductivity, the instability remains dynamic around
Mn atoms and the onset superconducting transition temperature keeps
nearly constant. These results indicate that the local lattice
instability may play an important role in the occurrence of HTSC.

\section{Acknowledgments}

The authors express their thanks to K. A. M\"{u}ller and A.
Bussmann-Holder for their valuable comments and discussions. The
EXAFS experiments were conducted under the proposals 2006G117,
2006G118 and 2007G071 at Photon Factory. This work was partially
supported by Japan Society for the Promotion of Science (JSPS).

\begin{table}
\caption{The average in-plane Cu-O bond distance and the M-O bond
distance (M=Mn, Ni, and Co) for the
La$_{1.85}$Sr$_{0.15}$Cu$_{1-x}$M$_x$O$_4$ (M=Mn, Ni, and Co) single
crystal samples taken at 300 K.}
\begin{ruledtabular}
\begin{tabular}{c c c}
sample composition & in-plane Cu-O bond distance (\AA) &
in-plane M-O bond distance (\AA) \\
 \hline La$_{1.85}$Sr$_{0.15}$CuO$_4$ & 1.885(8) & -
\\ Mn doped, $x$=0.025 & 1.886(2) & -
\\  Mn doped, $x$=0.03 & 1.890(1) & -
\\ Mn doped, $x$=0.05 & 1.892(2) & Mn-O: 1.918(4)
\\ Ni doped, $x$=0.03 & 1.886(8) & -
\\ Ni doped, $x$=0.05 & 1.886(3) & Ni-O: 1.885(6)
\\ Co doped, $x$=0.05 & 1.887(4) & Co-O: 1.888(2)
\\

\end{tabular}
\end{ruledtabular}
\end{table}

\end{document}